\documentclass[preprint]{aastex}
\usepackage{multicol}
\usepackage{natbib}
\newcommand{\kms}{km~s$^{-1}$}

\shorttitle{Kinematics of Open Clusters}
\shortauthors{Hayes & Friel}

\begin{document}

\title{RADIAL VELOCITIES OF THREE POORLY STUDIED CLUSTERS AND THE KINEMATICS OF OPEN CLUSTERS}
\author{Christian R. Hayes and Eileen D. Friel}
\affil{Astronomy Department, Indiana University, Bloomington, IN 47405, USA}
\email{hayescr@indiana.edu, efriel@indiana.edu}

\begin{abstract}
We present radial velocities for stars in the field of the open star clusters Berkeley 44, Berkeley 81, and NGC 6802 from spectra obtained using the Wisconsin--Indiana--Yale--NOAO (WIYN) 3.5 m telescope.  These clusters are of intermediate age (1-3 Gyr), located within the solar Galactocentric radius, and have no previous radial velocity measurements.  We find mean radial velocities of $-$9.6 $\pm$ 3.0 km s$^{-1}$, 48.1 $\pm$ 2.0 km s$^{-1}$, and 12.4 $\pm$ 2.8 km s$^{-1}$  for Be 44, Be 81, and NGC 6802, respectively.  We present an analysis of radial velocities of 134 open clusters of a wide range of ages using data obtained in this study and the literature.  Assuming the system of clusters rotates about the Galactic center with a constant velocity, we find older clusters exhibit a slower rotation and larger line--of--sight (LOS) velocity dispersion than younger clusters.  The gradual decrease in rotational velocity of the cluster system with age is accompanied by a smooth increase in LOS velocity dispersion, which we interpret as the effect of heating on the open cluster system over time.
\end{abstract}

\keywords{Galaxy: kinematics and dynamics - open clusters and associations: individual (Be 44, Be 81, NGC 6802)}

\section{Introduction}
	As they orbit in the Galactic disk, open star clusters are significantly affected by both internal interactions between stars, and their external interaction with the Galaxy.  As open clusters age, they lose stars due to internal dynamical effects and mass segregation, which  operates on timescales  on the order of several Gyr for an average cluster \citep{spitzer58}.  Compounding these effects of two-body relaxation, those due to interactions with the Galactic potential, giant molecular clouds and other massive objects may further accelerate cluster dissolution.  Since old open clusters can reach ages up to about 10 Gyr, yet are associated with the Galactic disk, some mechanism must have allowed them to survive the numerous interactions with the disruptive forces in the Galaxy.  Additionally, the oldest clusters are found to lag solar rotation, with higher line of sight (LOS) dispersions than young clusters of the thin disk \citep{s95,f02}.

This has led to questions of whether these old clusters have survived for so long because their initial velocities have taken them out of reach of the disruptive mechanisms, or if they acquired their current velocities through heating processes, yet were dense or massive enough to remain intact \citep{f13}.  It has also been suggested that old open clusters could be associated with structures other than the thin disk, such as the thick disk, the Sgr dwarf or other merger remnants. For example,  \citet{fr04}  suggested that clusters could be used (especially kinematically) to determine the nature of observed structures such as the Monoceros Ring in the Canis Major region.  Thus it is useful to study the kinematics of open clusters to determine their place in large-scale Galactic structure and evolution.  

To describe the motions of a single cluster however, individual stellar velocities of cluster members must be determined.  Since the stars in a cluster are born under the same conditions and are gravitationally bound, they will move through the Galaxy at a systematic velocity with small star-to-star variations due to internal velocity dispersions of about 1 km s$^{-1}$ \citep{f13}.  These small velocity dispersions provide a means for determining cluster membership even in the absence of proper motions.   Moreover, when combined with knowledge of the cluster color--magnitude diagram (CMD) and a star's evolutionary state, radial velocities can determine probable cluster members with relatively high precision. 

Combining cluster membership with photometry allows for more accurate determinations of properties such as age, distance, and to some extent metallicity.  Cluster membership is particularly important in crowded fields where field star contamination is high, because not only does it separate the cluster from the field, which allows for more accurate determination of the cluster properties, it also allows for specific selection of cluster stars for future observations with spectroscopy for elemental abundances. 

Motivated by all these considerations, we present new radial velocities and membership information for three relatively unstudied clusters, Berkeley 44, Berkeley 81 and NGC 6802, which have no previous radial velocity measures.  Prior to this study, only photometric data had been obtained for these clusters.  Berkeley 44 has been studied in the optical with $BVI$ photometry by \citet{car06} and by \citet{jh11} and in the IR using Two Micron All Sky Survey photometry by \citet{tur11}.  Cluster properties determined from these studies indicate Be 44 has an age of 1.3 -- 2.9 Gyr, reddening $E(B-V) = 0.98 - -1.40$, and intrinsic distance modulus of 11.26 -- 12.48, leading to a distance of 1.8 -- 3.1 kpc from the Sun.  We have chosen to utilize the most recent \citet{jh11} photometry for its combination of larger area of the sky, photometric depth, and secure photometric calibration.  Photometric properties of Be 81 have been determined by \citet{sg98} and most recently by  \citet{don14}.  These studies are in good agreement, indicating Be 81 has a distance of $\sim$ 3 kpc and an age of 0.7 -- 1.0 Gyr.  NGC 6802 has been studied photometrically by \citet{kal88} and \citet{net07} in searches for contact binaries and chemically peculiar stars, respectively, and by \citet{rica95} and \citet{jh11}, for the determination of cluster parameters.  These works agree in finding the cluster to be $\sim$ 1 Gyr in age, located 1.8 -- 2.0 kpc from the Sun.  

Adopted cluster parameters and sources for these properties are given in Table \ref{cl_summary}. The clusters are all located within the solar Galactocentric radius and are of intermediate age, making them interesting objects of study since few old clusters are found this far inside the solar Galactocentric radius.   It is especially interesting to study these clusters kinematically to determine whether they orbit near this radius, or have more elliptical orbits that take them to much larger radii.  

Additionally, we perform an analysis of the radial velocities of 134 clusters with properties obtained in this paper and from the literature.  These clusters exhibit a wide range in properties and locations, with ages from a few Myr to 10 Gyr, Galactocentric radii under 6 kpc to outside of 20 kpc, and distances from the Galactic plane of over 2 kpc.  With a sample of this size we are able to look at trends in kinematics with age and Galactic location.  We especially utilize the wide range in ages to learn more about the evolution of cluster rotation.
	
\section{Selection and Observation}

\subsection{Target Selections}

In general, our target selections were motivated by the desire to maximize the number of cluster members observed.  To make our target selections, we used the photometric data presented in \citet{jh11} for Be 44 and NGC 6802  and that recently gathered by \citet{don14} for Be 81, because these studies maximized the areal coverage and provided excellent photometric accuracy.  In general we selected stars in the cluster CMD within about 1 mag in $V$ and in $B-V$ around the red giant clump (RGC).  When possible, as for Be 44, we included brighter stars in the red giant region.  Most objects were selected to lie within twice the angular radius for each of the clusters as given in \citet{jh11} and  \citet{sg98}, but for Be 81, we were able to include a number of RGC stars outside of this region, in case the cluster extended farther than previously thought.   We also selected a few targets within two radii of the cluster center that appeared to be blue stragglers or in the main sequence turnoff region.  

\subsection{Observation and Data Reduction}

Our data were obtained using the Wisconsin-Indiana-Yale-NOAO (WIYN) 3.5 m telescope,\footnote{The WIYN Observatory is a joint facility of the University of Wisconsin-Madison, Indiana University, Yale University, and the National Optical Astronomy Observatory.} the Hydra multi-fiber positioner, the Bench Spectrograph, and a 2600 x 4000 pixel CCD (STA1) on the nights of  2012 May 7, 8, and 9  Spectra have a dispersion of 0.158 \AA \ pixel$^{-1}$ and cover a range of about 6082 -- 6397 \AA.  Typical  signal-to-noise ratio (S/N) were about 15--20 at the fainter $V$ magnitude limit of our sample, $V \sim16.5-17$ mag.  Four radial velocity standard stars were observed over the three nights of our observations with S/N \textgreater 100, along with stars in the fields of the Berkeley 44, Berkeley 81 and NGC 6802.

These data were reduced using standard IRAF\footnote{IRAF is distributed by the National Optical Astronomy Observatories, which are operated by the Association of Universities for Research in Astronomy, Inc., under cooperative agreement with the National Science Foundation.}  software to perform bias subtraction, flat fielding, dark subtraction, dispersion correction with ThAr lamp spectra (taken multiple times throughout each night to account for instrumental shifts), and sky subtraction.  In addition, the L.A.Cosmic algorithm \citep{vdo01} was used to remove cosmic rays, and multiple exposures were taken at each setup for a given cluster to improve our S/N and reduce any remaining effects of cosmic rays.

\section{Radial Velocities}

\subsection{Individual Stellar Velocities}

Radial velocities for individual stars were determined using the FXCOR task in the NOAO.RV package in IRAF, which performs a Fourier cross-correlation between an object spectrum and a template spectrum.  For each star, two to three different radial velocity measurements were obtained from cross-correlation with each of the templates (the velocity standards) taken on the same night.  To determine the uncertainty on the radial velocity measurements, each of the templates was cross-correlated with each other, and the measurements of the same template observed on different nights were cross-correlated with each other, to determine the template-to-template and night-to-night variations respectively.  However, both of these variations were on average around 0.15 km s$^{-1}$ with uncertainties on these values $\sim$ 0.5 km s$^{-1}$, consistent with zero.  On the other hand, uncertainties on the individual velocities as a result of the cross-correlation were generally under 1 km s$^{-1}$, yet typically several times the template-to-template and night-to-night variations.  Therefore, the uncertainty given for the individual radial velocity of a star is the average of the errors from the cross-correlation with each template.  In Table \ref{cl_vels}, we present new radial velocities for 132 stars in the fields of Be 44, NGC 6802, and Be 81.  In Table \ref{cl_vels}, the star ID, $V$ magnitude, and ($B-V$) color (Columns 1, 4, and 5) are those according to the \citet{jh11} photometry for Be 44 and NGC 6802, and according to the \citet{don14} photometry for Be 81.  The determined radial velocities and the uncertainties are given in Columns 2 and 3.  Membership status is given in Column 6, with explanation given in the following section.

\begin{figure}
\plotone{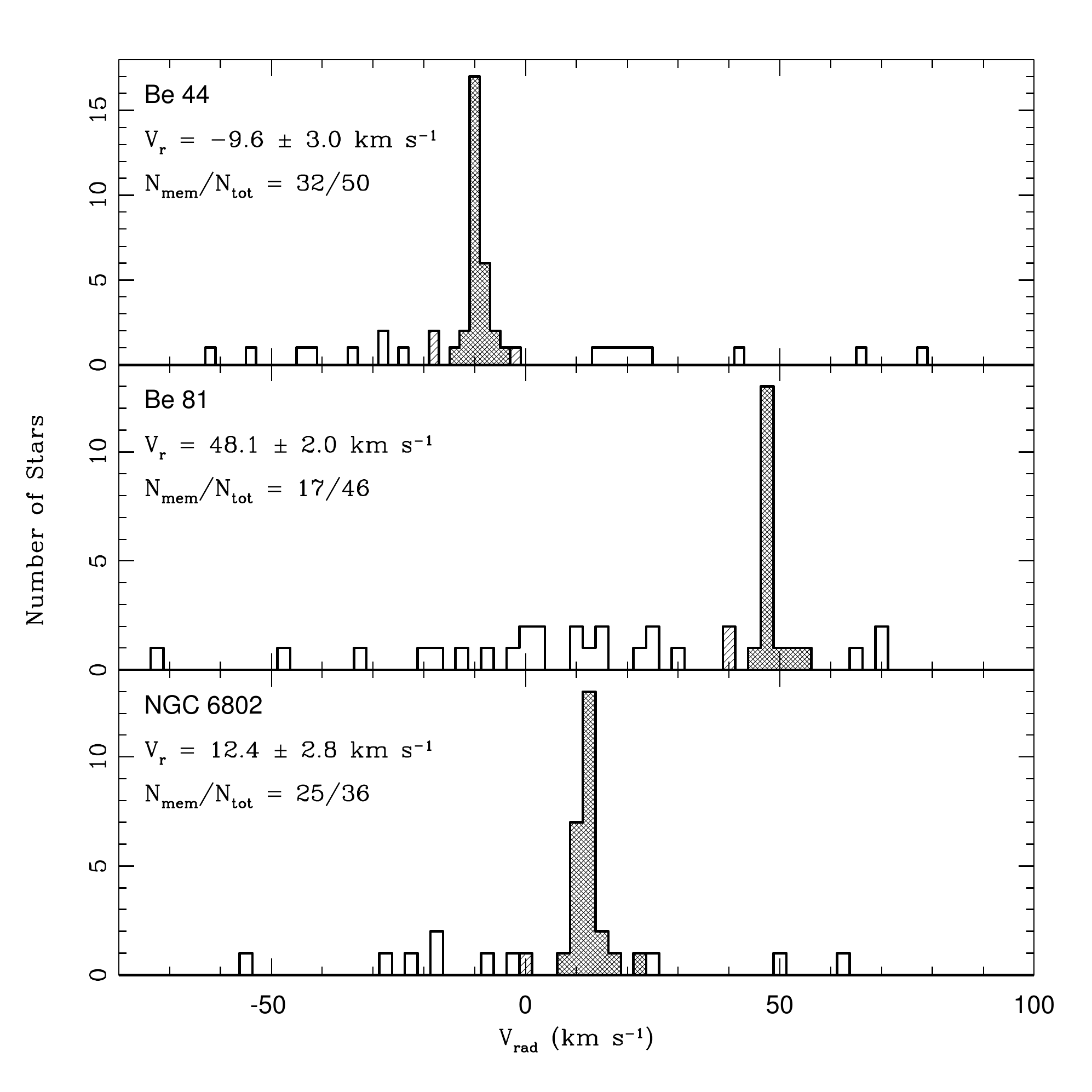}
\caption{Radial velocity histograms for Be 44, Be 81, and NGC 6802.  Cluster members are identified with crosshatching and stars with uncertain membership status are noted with hatched lines.}
\label{vr_hist}
\end{figure}

\subsection{Cluster Mean Velocities}

\begin{figure}
\plotone{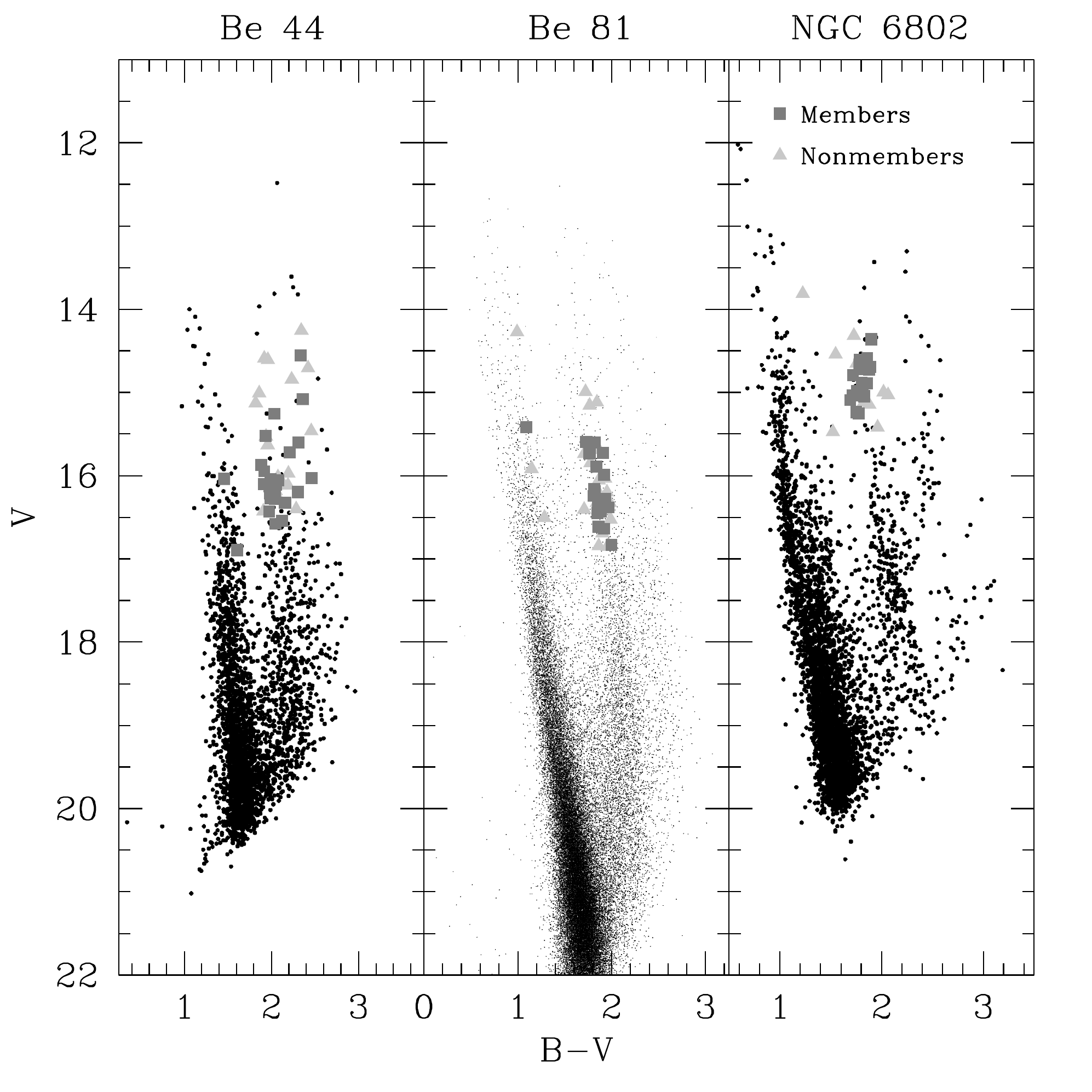}
\caption{CMDs for Be 44, Be 81, and NGC 6802, with observed cluster members given as gray squares and nonmembers as light gray triangles.  Note in Be 44, two bluer members, the brighter of which is likely a blue straggler, and the fainter, a turnoff star, were used in determination of the mean cluster velocity.  Note also the bluest member in Be 81, this is the supposed blue straggler, star 32081.}
\label{CMD}
\end{figure}

To compute mean cluster velocities, the possible cluster members must first be determined.  To do so, we examined the distributions of radial velocities in order to find a peak in the distribution of individual stellar velocities.  The generated histograms are shown in Figure \ref{vr_hist}, with the cluster members (or potential members) identified.  As the peaks in velocity are well defined for each cluster, we considered all stars more than 10 km s$^{-1}$ from the peak to be nonmembers. Any stars within about 5 km s$^{-1}$ on either side of the peak were defined as cluster members, since this range encompassed the primary distribution.  Those stars with radial velocities between 5 and 10 km s$^{-1}$ from the peak (noted with a \textquotedblleft \ ?\ \textquotedblright \ in Table \ref{cl_vels}), are potential members.  To decide whether or not to include these stars in the cluster mean velocities, we examined their position in the cluster CMD's and the star's radial distance from the cluster center.  For Be 44, we find three stars in this uncertain range, however they fall among other cluster members in both the CMD and position on the sky, so they are included in calculating the mean cluster velocity for Be 44.  For NGC 6802, only a single star (4650) falls in this range.  This star however was located at about 4 cluster radii from the center of the cluster.  Thus, we consider it unlikely that this star is member, and do not include it in the calculation of NGC 6802's mean velocity.  Finally, for Be 81, we find two stars of questionable membership status.  For star 53528 in Be 81 we find a situation similar to that of star 4650 in NGC 6802 (with 53528 also located at about 4 cluster radii from the cluster center), so we exclude star 53528 from our calculations.  For star 32081 in Be 81, we find a color and magnitude suggesting a potential blue straggler (if a true member of Be 81), which may explain its more deviant velocity.  However, even if star 32081 is a blue straggler (and likely in a binary), we would not resolve the velocity of the binary, so we exclude it from our mean cluster velocity calculations.  The final cluster members (including star 32081 in Be 81) are displayed in the CMDs for the clusters in Figure \ref{CMD}.

For Be 44, we found 32 members out of 50 stars giving a mean radial velocity of $-9.6 \pm 3.0$ km s$^{-1}$ (standard deviation).  For NGC 6802, we found 25 members out of 36 stars with a mean radial velocity of $12.4 \pm 2.8$ km s$^{-1}$.  For Be 81, 17 cluster members are found out of 46 stars giving a mean radial velocity of $48.1 \pm 2.0$ km s$^{-1}$.  Note that the proportion of members to nonmembers gives an estimate on the field contamination in each of these clusters.  We find a (pseudo) field contamination of 36\% in Be 44, about 30\% in NGC 6802, and about 63\% in Be 81.    As noted before, the observed stars were generally chosen to maximize the likelihood of finding cluster members, so this estimate is a lower limit to the actual field contamination.  These numbers emphasize the importance of determining cluster membership, as the large numbers of field stars will interfere with determining cluster properties. Radial velocity measures provide an important determination of cluster membership, which can then be used to clean the cluster CMDs of field star contamination and allow for accurate measures of cluster properties.  The determination of cluster properties based on these membership determinations will be the subject of later work, once cluster abundances are available.

\section{Analysis}

\subsection{Method}

For our analysis, we supplement our results with those for 131 clusters with measured ages, distances, and radial velocities.  The properties for these clusters were obtained from the literature and adopted values for all 134 clusters are given in Table \ref{sample_params}.  These clusters were chosen to sample a wide range of properties, particularly age and distance.   In the following analysis we assume a solar Galactocentric radius of 8.0 kpc, solar rotational velocity of 220 km s$^{-1}$, and a transformation of radial velocities to the local standard of rest (LSR) as given in \citet{bb99}.
	The method of analysis used here was developed by \citet{fw80}, who originally used it to examine the kinematics of globular clusters.  This method was also used later by \citet{z85} and \citet{a89} in their study of globular clusters.  The method defined by \citet{fw80} takes advantage of the geometric relation between the observed radial velocity of a cluster and its position, rotational velocity, expansion velocity, peculiar velocity, and the effect of solar rotation on the measured radial velocity.  
Here we neglect the expansion velocity term, as did \citet{fw80} and \citet{z85} who showed that the relationship simplifies \citep{a89}  to  

\begin{displaymath}
V_S = v_{rot} \cos(\psi)
\end{displaymath}
where $V_S$ is the radial velocity of a cluster with respect to a stationary observer at the Sun's location and 

\begin{displaymath}
V_S = V_{LSR} + v_{\odot} \cos(\lambda).
\end{displaymath}

Here, $V_{LSR}$ is the radial velocity of the cluster relative to the LSR, $v_{\odot}$ is the rotational velocity of the LSR about the Galactic center,   $\psi$ is the angle between the rotational-velocity vector of the cluster and the line passing through the cluster and the Sun, and $\lambda$ is the angle between the apex of the LSR and the position of the cluster on the sky.  \citep[provides a useful diagram of the geometry and definitions of  $\psi$ and $\lambda$  in terms of cluster position and distance.]{fw80}  Thus, assuming the entire cluster system rotates with a constant velocity, using a sample of clusters, one can determine a linear fit to the relation of $V_S$ and $\psi$, with the slope giving the rotational velocity of the system and the scatter around this fit providing the LOS dispersion in rotational velocities.  It should be noted that this LOS dispersion is a combination of effects of intrinsic rotational velocity dispersions and peculiar velocities, depending on the location of a given cluster in the Galaxy.

This method has also been applied to the old open cluster system.  \citet{f89} first examined a sample of 23 old (ages \textgreater \ 0.8 Gyr) open clusters using this method to find a rotational velocity of $203 \pm 9$ km s$^{-1}$ with a LOS dispersion of 22 km s$^{-1}$.  \citet{s95} also utilized this method, building upon this sample to find $211 \pm 7$ km s$^{-1}$ with a LOS dispersion of 28 km s$^{-1}$ for 35 old (ages \textgreater \ 1 Gyr) open clusters.  Given a significantly larger sample size, which includes young clusters as well as old, we are able to sub-divide our sample based on different properties to examine the rotational velocities and dispersions in these sub-samples.  Additionally, many of the clusters in our sample have more accurate radial velocities than in the previous studies, which had uncertainties on the order of $\sim 10$ km s$^{-1}$.   Most of the clusters in this sample have uncertainties in their radial velocities of $\sim 1$ km s$^{-1}$ or less.  

	 Table \ref{sample_params} gives the properties for the 134 clusters analyzed in this survey, along with the calculated values of $V_S$ and cos($\psi$) (given in Columns 10 and 11).  The cluster mean radial velocity and standard error of the mean given in Columns 4 and 5 were obtained from a variety of sources, which are referenced in the last column.  True distance moduli and ages (log(Age) given in Column 9) were obtained from a variety of photometric studies of these clusters, also referenced in the last column.

\subsection{Rotational Velocities of the Open Cluster System}

\begin{figure}
\plotone{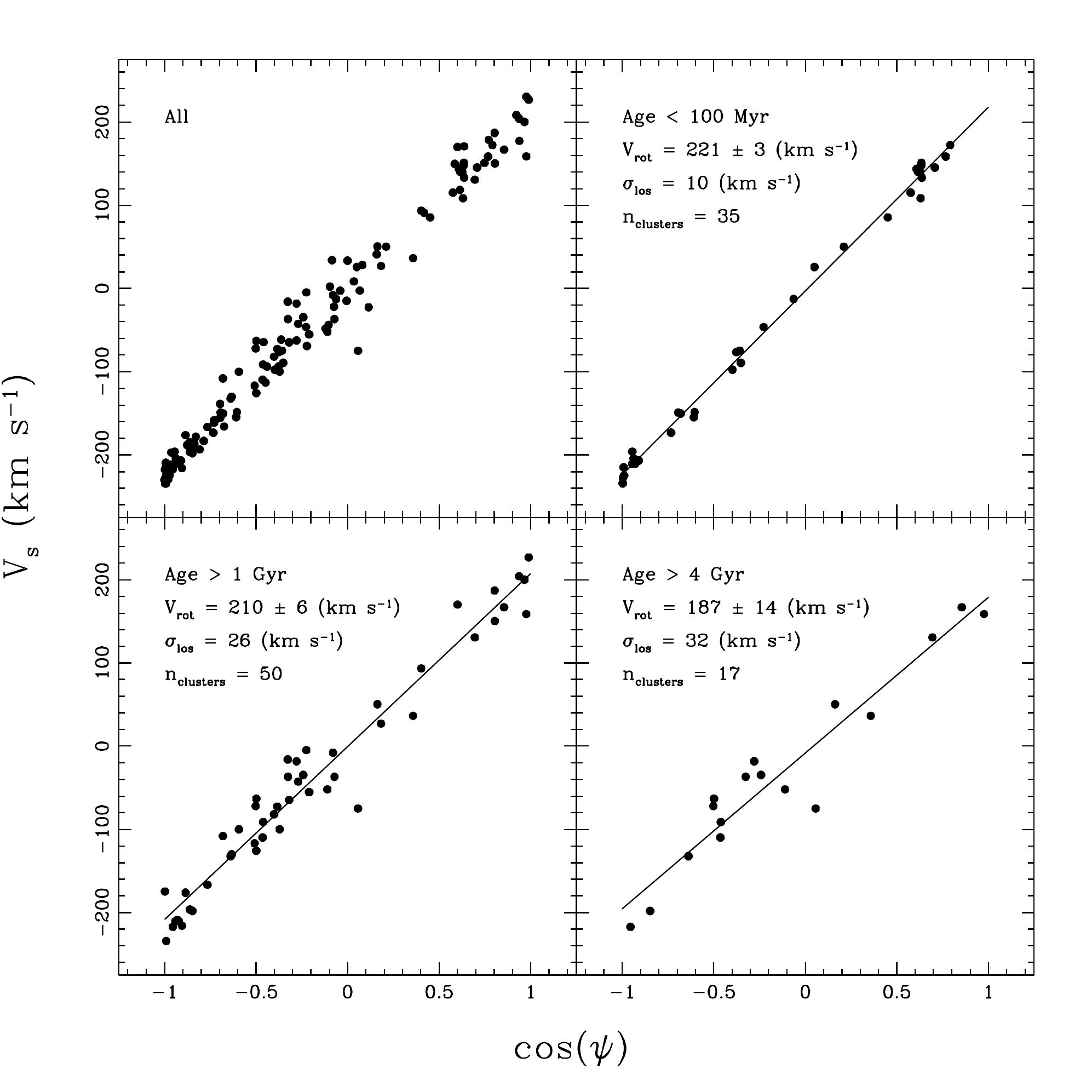}
\caption{Plots of $V_{S} = v_{rot} \cos(\psi)$ for clusters in the sample, binned by age.  Top left: the relationship for all clusters in the sample.  Top right: young clusters with ages under 100 Myr.  Bottom left: clusters with ages over 1 Gyr.  Bottom right: clusters older than 4 Gyr.  The solid lines are linear fits to the points; resulting parameters are given in each panel.}
\label{vs_rot}
\end{figure}

In Figure \ref{vs_rot}, we plot $V_S$ and cos($\psi$) for four sets of data, the entire sample of 134 clusters, young clusters with ages under 100 Myr, old clusters with ages over 1 Gyr, and the oldest of those clusters with ages over 4 Gyr.  For the three subsets presented in Figure \ref{vs_rot}, we have fit a line to the data, determining the rotational velocity of the system of clusters  and the LOS velocity dispersion for the same.  While individual outliers are present, in general, the younger subset follows a tight distribution and steeper slope (larger rotational velocity) around the best fit line, whereas the older subsets exhibit more scatter around a more shallow slope.  We shall examine these in further detail.

First, we examine the system of young clusters with ages under 100 Myr.  We have a sample of 35 clusters which give a rotational velocity of $221 \pm 3$ km s$^{-1}$ and a LOS dispersion of 10 km s$^{-1}$.  This is consistent with the local rotational velocity of $229 \pm 18$ km s$^{-1}$ \citep{g08}, implying that young clusters are, not surprisingly, members of the thin disk, where they recently formed.  

In an effort to compare to previous results, we present the results of this analysis applied to the old open clusters, those with ages over 1 Gyr.  In this age range, we improve the sample size from 35 in \citet{s95} to 50 clusters, and find a rotational velocity of $210 \pm 6$ km s$^{-1}$ with a LOS dispersion of 26 km s$^{-1}$.  This result is consistent with both \citet{s95} and \citet{f89}, in both the rotational velocity and dispersion of the old open cluster system.

As the sample of older (\textgreater \, 1 Gyr) clusters is generally weighted toward 1 Gyr old clusters, it is of interest to examine the oldest of these clusters.  We have 17 clusters with ages greater than 4 Gyr in age, increasing by 50\% the sample of 12 given by \citet{s95}.  Note that this implies that about two thirds of our old clusters have ages 1-4 Gyr, so if any interesting motions existed only in the oldest clusters, they would likely be overshadowed by the motions of the relatively younger clusters in our previous range.  However, with such a small sample, individual outliers will also have more influence on our results.  With both of these ideas in mind, for the system of the oldest clusters, we find a rotational velocity of $187 \pm 14$ km s$^{-1}$ and a LOS dispersion of 32 km s$^{-1}$.  While this is a significantly lower rotational velocity than that found for clusters older than 1 Gyr, we suspect that is largely due to the small sample size and the thus increased influence of two clusters, Berkeley 17 and NGC 6791, which will be examined in more detail in Section 5.  Note that the fits for each of the three age ranges examined above are given in Table \ref{vr_solns}.  

\begin{figure}
\plotone{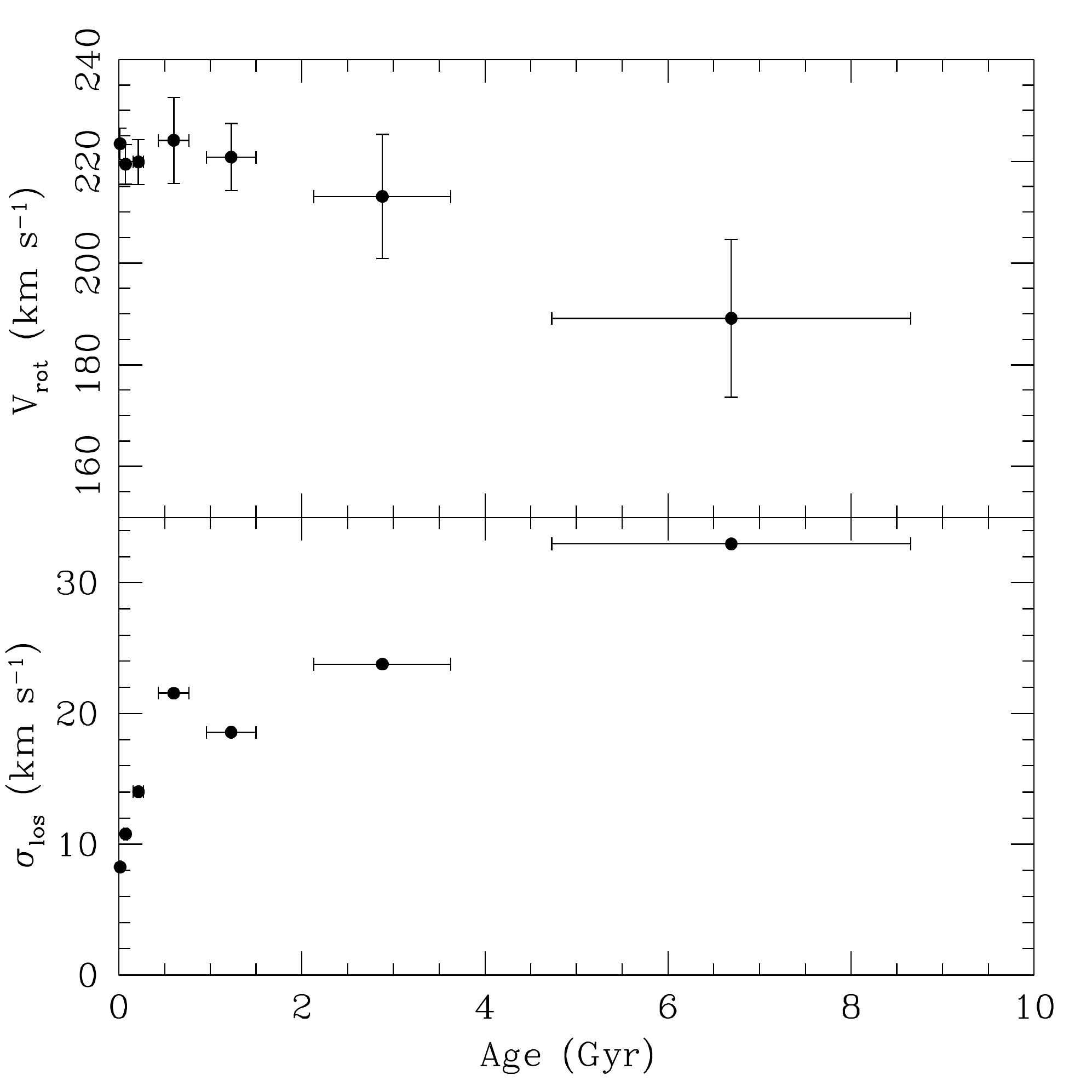}
\caption{Plots of age vs. rotational velocity (top) and age vs. line--of--sight velocity dispersion (bottom) of the seven age groups as described in the text and Table \ref{vr_solns}.  
}
\label{vs_age}
\end{figure}

As the old cluster system exhibits a lower rotation velocity and a significantly higher LOS velocity dispersion than the young cluster system, it is of interest to examine the distribution of these properties as a function of age in more detail, which our enlarged sample size allows.  Thus we separate our sample into seven age bins (each with about the same number of clusters) and apply the same analysis as above, the results of which are given in Table \ref{vr_solns}. Figure \ref{vs_age} plots these LOS velocity dispersions and rotational velocities against the average age of each age group.  As ages generally have high uncertainties, we examined the effects of changing the definition and ranges of these age groups, to account for the possibility of incorrect ages.  However, we found that while specific values for LOS dispersion and rotational velocity changed slightly (largely due to changing the number of clusters in the range), the general trends were conserved.  Therefore we conclude that the trends shown in Figure \ref{vs_age} are real, and not the result of the particular age range selections we have made.

\section{Discussion}

\subsection{Age-Velocity Relationship}

The most striking feature in Figure \ref{vs_age} is that with the exception of the point at an age range around 600 Myr, there seems to be a smooth increase in LOS velocity dispersion and a smooth decrease in rotational velocity with age.  Additionally, the deviation of the 600 Myr point is due to two clusters (which will be examined later), and not due to a larger scatter amongst all of the clusters in this age range.  Thus by excluding the two deviant clusters, the 600 Myr point falls along the relationship established by the other age ranges.
Rather than displaying discontinuities or significant jumps or variations that might be interpreted in terms of particular dynamical interactions or specific events or individual subpopulations within the cluster system, it appears that the increase in the LOS dispersions accompanied by the decrease in rotational velocity occurs gradually over time for clusters.  This indicates that as clusters age, they transfer rotational velocity into random motions.  As further evidence of this, the increase in LOS dispersion and decrease in rotational velocity with increasing age are consistent in both shape and magnitude with those found in the Geneva--Copenhagen Survey (GCS) of the solar neighborhood based on the full $U$, $V$, and $W$ velocities of thin disk stars \citep{n04,h07,h09}.  The GCS studies found smooth increases of $U$, $V$, and $W$ velocity dispersions with age in thin disk stars, concluding that this reflects the effects of heating processes in the disk (as a combination of ÒpureÓ or random heating and interactions with large-scale structures: \citealt{h07}).  Since most of the clusters in our sample are in or close to the plane of the Galaxy, the $W$ velocities do not contribute much to the observed radial velocities.  However, our radial velocities are much more sensitive to $U$ and $V$ velocities, with $V$ velocities measured to an extent by the rotational velocity and the $U$ and $V$ velocity dispersions combined reflected in the LOS dispersion.  For the age ranges covered in the GCS studies, our LOS dispersion is consistent with the $U$ and $V$ field star velocity dispersions.  For example, in the oldest cluster group we examined, we found an LOS dispersion of 33 km s$^{-1}$ which falls between the $U$ velocity dispersion of about 40 km s$^{-1}$ and the $V$ velocity dispersion of about 23 km s$^{-1}$ found in \citet{h07} for field stars in the same age range.  Thus we find that the open clusters in our sample (all 134 clusters) are consistent with being thin disk objects and affected by heating processes, in a largely similar way as are individual stars.  

This increase in velocity dispersion for clusters had been found previously, but from a cluster sample much more limited in age and distance.  In a study using proper motions  \citet{wu09} determined $U$, $V$, and $W$ velocity dispersions for clusters in three age ranges (those under 500 Myr, those between 500 Myr and 1 Gyr, and those between 1 and 2 Gyr).  They found that velocity dispersions also increased with age, similar to our LOS dispersions, and as noted by \citet{wu09} reflect the stellar velocity dispersions found in the GCS studies.  The sample analyzed here extends this result to significantly larger ages and extent in the Galactic disk, finding results from radial velocities that are fully consistent with those using proper motions and full space velocities.

\subsection{Unusual Clusters}

We noted before that two of the oldest clusters, Berkeley 17 (8 Gyr, $V_S$ = 50.4 km s$^{-1}$, cos($\psi$) = 0.163) and NGC 6791 (10 Gyr, $V_S$ = 158.7 km s$^{-1}$, cos($\psi$) = 0.976),  deviate significantly from the bulk of other clusters in the $V_S$--cos($\psi$) diagram of all clusters (see Figure \ref{vs_rot}).  However, it is interesting to note that NGC 6791 does not appear to deviate much in the cluster group with ages greater than 4 Gyr, which may be significant or due to a lack of clusters in the same region on this diagram.  Be 17 is  located almost directly in the Galactic anti-center ($l = 175.65^{\circ}$), yet has a highly negative radial velocity of $-84$ km s$^{-1}$, implying a substantial $U$ velocity.  The kinematics of NGC 6791, on the other hand, have been studied in more detail to determine orbital parameters.  \citet{b06} found that NGC 6791 orbits with an eccentricity of about 0.5 taking it to a perigalacticon of about 3 kpc and a maximum height above the plane of about 1 kpc, all of which are consistent with having an eccentric thin disk orbit, which is interesting since NGC 6791 is one of the oldest open clusters known.  This is noted by \citet{b06}, who come to the conclusion that the most likely reason for NGC 6791's longevity is its high density and mass.  Indeed, \citet{pla11} finds a lower limit to the mass of approximately 5000 $M_\odot$ for NGC 6791, placing it among the most massive of known open clusters.  Additionally,  NGC 6791 will have suffered significant mass loss over its 8--10 Gyr lifetime \citep{jil12}, so would have been appreciably more massive in the past.  When considering the effect that heating would have over the lifetime of these clusters (8 - 10 Gyr), it is not unexpected that a number of clusters would have more eccentric orbits, explaining some of the kinematic properties of the oldest cluster system.  However, clusters will also share the motions of the clouds from which they formed, with the possibility that their somewhat more eccentric orbits are at least partly primordial.  Additionally, clusters could interact with a giant molecular cloud (or other massive body) in a particular way that imparts a large amount of energy to the cluster without disrupting it, thus leaving the cluster on a more eccentric orbit.  Thus, there are other methods of setting clusters on eccentric orbits which could also be in effect for these clusters.
	
As previously noted, the 600 Myr group of clusters does not appear to follow the smooth increase in LOS dispersion established by the rest of the points.  This appears to be due to two clusters with deviant velocities, NGC 1817 ($V_S$ = 33.9 km s$^{-1}$, cos($\psi$) = $-$0.086) and NGC 1912 ($V_S$ = $-$22.7 km s$^{-1}$, cos($\psi$) = 0.115), since excluding them from the group, both the $V_{rot}$ and LOS dispersion fall along the smooth curve defined by the rest of the age groups.  These two clusters are both located near the Galactic anti-center with NGC 1817 at $l = 186.16^{\circ}$ and NGC 1912 at $l = 175.25^{\circ}$, with radial velocities of 65 km s$^{-1}$ and $-45$ km s$^{-1}$, respectively.  Both of these appear significant and point to more eccentric orbits.  Given that these clusters have ages of 400--500 Myr, it appears they may have formed with more eccentric motions or have (as mentioned above) interacted with some structure in the Galaxy in a particular manner that set these clusters on more eccentric orbits.

\section{Conclusion}

We have derived radial velocities for 3  open clusters with no previous velocity measurements: Be 44, Be 81, and NGC 6802.  Combining these velocities with results from the literature to assemble a sample of 134 clusters of a wide range of ages and distances, we investigate the kinematics of the open cluster system.   We find that the youngest clusters, with ages less than 100 Myr, rotate with a velocity of $221 \pm 3$ \kms~ with a LOS velocity dispersion of only 10 \kms, as expected for a young disk population.   However, older clusters show systematically slower rotational velocities and higher LOS velocity dispersions, which suggests that as clusters age, a portion of their rotational velocity is dispersed into random motions.   Additionally, the increase in velocity dispersion is consistent in both shape and magnitude with the velocity dispersions of local field stars as determined in the GCS.  This appears to indicate that clusters are affected by heating as individual stars are, and that these clusters are thin disk objects.  

While this study improved upon the sample size and quality of data of past studies of its kind, there are a number of ways in which this work can be expanded and improved to better understand open cluster kinematics.  There remain many more known open clusters that either do not have well defined radial velocities or lack them entirely.  Similarly, there are many clusters that lack detailed properties in general, such as age or distance.  By obtaining or improving the quality of these properties, the sample here could be greatly increased, allowing finer details of the age--rotation velocity and age--velocity dispersion relationships to be probed.  It would also be interesting to examine the effects of dividing clusters into metallicity ranges and applying the method of analysis used here, since many of the clusters in this sample lack metallicity measurements.  Also, while a number of clusters have measured proper motions and full space velocities derived from them, these measurements are often limited to the brightest magnitudes, so sample a limited number of  stars in the cluster field or do not reach very great distances.   Thus by improving and obtaining proper motion measurements and full space velocities, cluster kinematics could be compared directly with Galactic field star kinematics in terms of well defined $U$, $V$, and $W$ velocities and velocity dispersions.  There are many directions that can be taken to learn more about open clusters and their interaction with the Galaxy.

\acknowledgments
This research has made use of the WEBDA database, operated at the Institute for Astronomy of the University of Vienna.  C.R. Hayes thanks the Cox Research Scholars Program at Indiana University for making this research possible.  Thanks to Enrico Vesperini and Anna Lisa Varri for the helpful discussion about our results, and to Paolo Donati for early access to his photometry of Berkeley 81.

\begin{deluxetable}{l c c c c c c r c}
\tablewidth{0pt}
\tablecolumns{9}
\tablecaption{Properties of Observed Clusters}
\tablehead{\colhead{} & \colhead{} & \colhead{} & \colhead{} & \colhead{} & \colhead{$d$} & \colhead{$R_{gc}$} & \colhead{$z$} & \colhead{Age} \\ \colhead{Cluster} & \colhead{$l$} & \colhead{$b$} & \colhead{$(m-M)_{o}$} & \colhead{$E(B-V)$} & \colhead{(pc)} & \colhead{(kpc)} & \colhead{(kpc)} & \colhead{(Gyr)}}
\startdata
Be 44\tablenotemark{a} & 53.21 & 3.35 & 12.5 & 0.98 & 3100 & 6.6 & 0.18 & 2.9\\
Be 81\tablenotemark{b} & 34.51 & $-$2.07 & 12.6 & 1.0 & 3200 & 5.6 & $-$0.12 & 1.0\\
NGC 6802\tablenotemark{a} & 55.32 &  0.92 & 11.3 & 0.84 & 1800 & 7.1 & 0.03 & 1.0\\
\enddata
\tablenotetext{a}{Photometric properties obtained by \citet{jh11}.}
\tablenotetext{b}{Photometric properties obtained by \citet{sg98}.}
\label{cl_summary}
\end{deluxetable}

\begin{deluxetable}{lrcccclrcccc}
\tabletypesize{\scriptsize}
\tablewidth{0pt}
\tablecolumns{12}
\tablecaption{Individual Stellar Velocities and Membership}
\tablehead{\colhead{ID} & \colhead{$V_r$} & \colhead{$\epsilon$} & \colhead{$V$} & \colhead{$B-V$} & \colhead{Member?} & \colhead{ID} & \colhead{$V_r$} & \colhead{$\epsilon$} & \colhead{$V$} & \colhead{$B-V$} & \colhead{Member?} \\ 
\colhead{} & \colhead{(km s$^{-1}$)} & \colhead{(km s$^{-1}$)} & \colhead{} & \colhead{} & \colhead{} & \colhead{} & \colhead{(km s$^{-1}$)} & \colhead{(km s$^{-1}$)} & \colhead{} & \colhead{} & \colhead{}}
\startdata
\cutinhead{Berkeley 44\tablenotemark{a}}
  219 & 14.6 & 1.7 & 16.42 & 1.91 & No & 5214 & $-$9.5 & 0.9 & 16.27 & 1.99 & Yes \\
  689 & $-$43.8 & 0.7 & 15.01 & 1.86 & No & 5467 & $-$9.8 & 0.5 & 15.72 & 2.21 & Yes \\
  799 & 16.4 & 0.6 & 14.26 & 2.35 & No & 5602 & $-$8.6 & 1.1 & 16.22 & 2.01 & Yes \\
  1212 & 66.0 & 1.2 & 16.25 & 2.07 & No & 5623 & $-$9.5 & 0.8 & 15.08 & 2.36 & Yes \\
  1800 & 20.3 & 0.9 & 16.12 & 2.14 & No & 5694 & $-$9.3 & 0.9 & 16.54 & 2.12 & Yes \\
  1987 & 23.3 & 0.8 & 16.18 & 2.05 & No & 5925 & $-$8.7 & 0.7 & 15.95 & 1.92 & Yes \\
  2310 & 78.2 & 0.9 & 15.46 & 2.46 & No & 6129 & $-$9.7 & 1.4 & 16.21 & 2.03 & Yes \\
  2446 & $-$8.5 & 1.0 & 15.87 & 1.88 & Yes & 6381 & 22.2 & 1.0 & 15.13 & 1.82 & No \\
  2498 & $-$9.2 & 1.6 & 16.32 & 2.16 & Yes & 6472 & $-$9.6 & 0.8 & 16.06 & 2.08 & Yes \\
  2624 & $-$23.6 & 2.2 & 15.53 & 1.94 & No & 6756 & $-$8.8 & 0.7 & 16.14 & 2.0 & Yes \\
  3299 & $-$28.0 & 0.8 & 15.64 & 1.96 & No & 6860 & $-$62.9 & 0.6 & 14.59 & 1.92 & No \\
  3391 & $-$34.1 & 1.1 & 16.40 & 2.29 & No & 6959 & $-$4.5 & 0.7 & 16.12 & 2.04 & Yes \\
  3589 & $-$9.3 & 0.8 & 15.52 & 1.93 & Yes & 7339 & $-$10.1 & 0.7 & 16.22 & 1.98 & Yes \\
  3635 & 18.4 & 1.2 & 14.84 & 2.23 & No & 7519 & $-$9.6 & 1.9 & 16.13 & 1.98 & Yes \\
  3716 & $-$5.2 & 0.5 & 16.21 & 2.02 & Yes & 7621 & $-$9.2 & 1.1 & 16.01 & 1.91 & Yes \\
  4025 & $-$18.4 & 0.5 & 15.60 & 2.31 & Yes? & 7785 & $-$9.5 & 0.6 & 16.28 & 2.04 & Yes \\
  4126 & $-$6.5 & 0.6 & 14.55 & 2.34 & Yes & 7873 & $-$8.3 & 1.4 & 16.04 & 1.46 & Yes \\
  4159 & $-$8.4 & 0.8 & 16.20 & 2.02 & Yes & 8677 & $-$9.1 & 0.9 & 16.89 & 1.61 & Yes \\
  4284 & $-$9.3 & 1.0 & 16.43 & 1.98 & Yes& 8803 & $-$9.6 & 0.8 & 16.11 & 2.06 & Yes \\
  4556 & $-$28.6 & 0.8 & 16.11 & 2.19 & No & 8857 & $-$18.3 & 0.8 & 15.26 & 2.04 & Yes? \\
  4672 & $-$10.6 & 0.7 & 16.58 & 2.05 & Yes & 8985 & $-$10.2 & 1.3 & 16.04 & 1.99 & Yes \\
  4682 & 41.8 & 1.1 & 15.98 & 2.19 & No & 9272 & $-$73.6 & 0.8 & 14.60 & 1.96 & No \\
  4922 & $-$11.1 & 1.4 & 16.20 & 2.04 & Yes & 9346 & $-$12.3 & 0.7 & 16.18 & 2.0 & Yes \\
  4946 & $-$13.7 & 0.6 & 16.20 & 2.3 & Yes & 10344 & $-$41.2 & 0.9 & 14.70 & 2.42 & No \\
  5141 & $-$3.0 & 0.9 & 16.03 & 2.46 & Yes? & 10913 & $-$53.9 & 0.9 & 16.01 & 2.08 & No \\
  \tablebreak
  \cutinhead{Berkeley 81\tablenotemark{b}}
  240 & 22.6 & 1.3 & 16.40 & 1.88 & No & 32887 & $-$17.9 & 1.9 & 14.27 & 0.99 & No\\
  648 & 2.3 & 0.8 & 16.06 & 1.86 & No & 32890 & 9.6 & 1.0 & 15.15 & 1.76 & No\\
  4820 & 26.1 & 0.7 & 16.31 & 1.88 & No & 32908 & 47.2 & 0.6 & 15.60 & 1.82 & Yes\\
  5406 & 46.8 & 1.9 & 16.64 & 1.92 & Yes & 32909 & $-$18.9 & 1.5 & 15.68 & 1.82 & No\\
  11569 & 1.4 & 0.7 & 16.32 & 1.84 & No & 32910 & 25.3 & 0.8 & 15.69 & 1.86 & No\\
  15220 & 93.5 & 1.7 & 16.52 & 1.99 & No & 32916 & 0.1 & 1.1 & 15.85 & 1.78 & No\\
  16463 & 46.0 & 1.1 & 16.61 & 1.86 & Yes & 32918 & 47.1 & 0.7 & 15.99 & 1.92 & Yes\\
  17858 & 46.9 & 1.1 & 16.83 & 2.00 & Yes & 32928 & 70.9 & 0.9 & 16.03 & 1.93 & No\\
  24615 & 48.2 & 0.8 & 16.24 & 1.81 & Yes & 32931 & 14.9 & 1.0 & 16.20 & 1.95 & No\\
  24728 & 49.1 & 1.6 & 16.36 & 1.85 & Yes & 32941 & $-$3.2 & 1.1 & 16.25 & 1.93 & No\\
  24956 & 70.4 & 1.1 & 16.85 & 1.95 & No & 32944 & 47.5 & 1.2 & 16.38 & 1.96 & Yes\\
  25710 & 65.4 & 2.0 & 16.85 & 1.97 & No & 32945 & 9.9 & 1.2 & 16.36 & 1.95 & No\\
  25830 & 48.2 & 1.0 & 16.43 & 1.88 & Yes & 32952 & 89.7 & 0.9 & 16.42 & 1.90 & No\\
  27396 & 12.3 & 1.1 & 16.69 & 1.90 & No & 32963 & 29.4 & 1.9 & 16.41 & 1.71 & No\\
  30656 & $-$72.5 & 2.6 & 16.31 & 1.88 & No & 32971 & $-$31.4 & 1.4 & 15.92 & 1.15 & No\\
  32076 & $-$6.5 & 0.6 & 15.74 & 1.71 & No & 35205 & 15.0 & 0.6 & 14.99 & 1.72 & No\\
  32079 & 46.8 & 1.1 & 16.16 & 1.81 & Yes & 35255 & 0.9 & 1.9 & 16.50 & 1.28 & No\\
  32081 & 39.0 & 1.6 & 15.42 & 1.09 & No? & 53528 & 38.9 & 0.8 & 16.33 & 1.98 & No?\\
  32085 & 47.8 & 0.7 & 15.73 & 1.90 & Yes & 58063 & 51.6 & 0.8 & 15.59 & 1.73 & Yes\\
  32087 & 47.2 & 0.7 & 16.28 & 1.93 & Yes & 58067 & 46.6 & 1.3 & 15.73 & 1.76 & Yes\\
  32088 & 48.1 & 1.3 & 16.19 & 1.82 & Yes & 58107 & $-$13.7 & 1.0 & 16.85 & 1.86 & No\\
  32091 & 47.9 & 1.2 & 16.45 & 1.85 & Yes & 79900 & 33.4 & 1.3 & 16.19 & 1.85 & No\\
  32886 & $-$47.5 & 1.3 & 15.12 & 1.84 & No & 46455-u\tablenotemark{c} & 54.5 & 1.0 & 15.89 & 1.84 & Yes\\
  \tablebreak
  \cutinhead{NGC 6802\tablenotemark{a}}
  2167 & 14.6 & 0.8 & 14.89 & 1.86 & Yes & 7107 & 49.5 & 0.5 & 15.0 & 2.02 & No \\
  2467 & $-$3.3 & 0.6 & 15.14 & 1.88 & No & 7314 & 15.0 & 1.6 & 15.04 & 1.83 & Yes \\
  2519 & $-$18.3 & 0.6 & 15.03 & 2.06 & No & 7496 & 7.3 & 0.7 & 15.25 & 1.77 & Yes \\
  2705 & 9.1 & 0.9 & 15.09 & 1.69 & Yes & 7577 & 11.9 & 0.8 & 14.79 & 1.72 & Yes \\
  4052 & 10.9 & 0.9 & 14.98 & 1.76 & Yes & 7719 & 12.6 & 0.7 & 14.53 & 1.8 & Yes \\
  4608 & $-$27.9 & 1.4 & 13.81 & 1.22 & No & 7753 & 12.5 & 0.7 & 14.85 & 1.72 & Yes \\
  4650 & 1.0 & 0.5 & 14.31 & 1.73 & No? & 7783 & 22.6 & 0.7 & 14.14 & 1.79 & Yes \\
  5044 & 9.7 & 0.5 & 14.73 & 1.88 & Yes & 7967 & $-$18.7 & 0.9 & 14.65 & 1.76 & No \\
  5275 & 11.9 & 0.6 & 14.59 & 1.86 & Yes & 8271 & 11.1 & 0.7 & 14.92 & 1.75 & Yes \\
  5485 & 11.0 & 0.7 & 14.96 & 1.81 & Yes & 8346 & 13.0 & 0.7 & 15.45 & 1.86 & Yes \\
  5548 & 11.1 & 0.7 & 14.72 & 1.82 & Yes & 8538 & 16.8 & 1.7 & 15.39 & 1.84 & Yes \\
  5821 & 12.0 & 0.6 & 14.88 & 1.8 & Yes & 9056 & 13.0 & 1.0 & 15.04 & 1.81 & Yes \\
  5893 & $-$7.3 & 0.9 & 14.54 & 1.55 & No & 9348 & 61.7 & 0.7 & 15.03 & 1.8 & No \\
  6139 & 11.6 & 0.5 & 14.36 & 1.9 & Yes & 10173 & 25.2 & 2.0 & 15.48 & 1.52 & No \\
  6358 & 13.6 & 0.8 & 14.69 & 1.89 & Yes & 10183 & 12.1 & 0.8 & 14.72 & 1.82 & Yes \\
  6689 & 13.0 & 0.7 & 14.6 & 1.79 & Yes & 12472 & $-$23.2 & 1.3 & 15.42 & 1.96 & No \\
  6805 & 12.0 & 0.9 & 15.24 & 1.75 & Yes & 12529 & 11.2 & 0.8 & 15.04 & 1.71 & Yes \\
  6883 & 11.4 & 1.1 & 14.67 & 1.78 & Yes & 12587 & $-$55.5 & 0.5 & 15.12 & 1.83 & No \\
\enddata
\tablenotetext{a}{Star ID and photometry from \citet{jh11}.}
\tablenotetext{b}{Star ID and photometry from \citet{don14}.}
\tablenotetext{c}{Star ID and photometry unpublished, from uncut photometry of \citet{don14} received in private communication (star ID in uncut photometry is 46455).}
\label{cl_vels}
\end{deluxetable}

\begin{deluxetable}{l r r r c c c l c r c l}
\tabletypesize{\scriptsize}
\tablewidth{0pt}
\tablecolumns{12}
\tablecaption{Cluster Properties and Mean Radial Velocities}
\tablehead{\colhead{Cluster} & \colhead{$l$} & \colhead{$b$} & \colhead{$V_r$} & \colhead{$\epsilon$} & \colhead{Dist} & \colhead{$R_{gc}$} & \colhead{$z$} & \colhead{log(Age)} & \colhead{$V_S$\tablenotemark{a}} & \colhead{cos$(\psi)$\tablenotemark{a}} & \colhead{Reference ID} \\ 
\colhead{} & \colhead{} & \colhead{} & \colhead{(km s$^{-1}$)} & \colhead{(km s$^{-1}$)} & \colhead{(pc)} & \colhead{(kpc)} & \colhead{(kpc)} & \colhead{} & \colhead{(km s$^{-1}$)} & \colhead{} & \colhead{}}
\startdata
  Be 17 & 175.65 & -3.65 & -84.0 & 3.0 & 2800 & 10.7 & -0.175 & 9.9 & -74.9 & 0.056 & 18, 113\\
  Be 18 & 163.63 & 5.02 & -5.5 & 1.1 & 6000 & 13.8 & 0.52 & 9.6 & 50.4 & 0.163 & 67, 135\\
  Be 20 & 203.48 & -17.37 & 75.7 & 2.4 & 3200 & 10.9 & -0.97 & 9.8 & -18.3 & -0.279 & 40, 48\\
  Be 21 & 186.84 & -2.51 & -0.6 & 1.4 & 5200 & 13.2 & -0.23 & 9.4 & -36.8 & -0.072 & 125, 136\\
  Be 25 & 226.61 & -9.69 & 134.3 & 0.2 & 11300 & 17.6 & -1.9 & 9.7 & -36.8 & -0.326 & 24, 25\\
  Be 29 & 197.98 & 8.02 & 28.4 & 3.6 & 14500 & 22.1 & 2.017 & 9.6 & -52.0 & -0.111 & 48, 124\\
  Be 31 & 206.25 & 5.12 & 55.7 & 0.7 & 9200 & 16.8 & 0.825 & 9.3 & -55.2 & -0.21 & 57, 135\\
  Be 32 & 207.95 & 4.4 & 101.0 & 3.0 & 3800 & 11.4 & 0.289 & 9.5 & -15.9 & -0.327 & 57, 113\\
  Be 39 & 223.46 & 10.1 & 55.0 & 2.5 & 4600 & 11.7 & 0.801 & 9.8 & -109.8 & -0.464 & 48, 68\\
  Be 44 & 53.21 & 3.35 & 9.6 & 0.5 & 3100 & 6.6 & 0.183 & 9.46 & 200.1 & 0.966 & 63, 137\\
  Be 66 & 139.43 & 0.22 & -50.6 & 0.3 & 5800 & 12.9 & 0.022 & 9.5 & 93.4 & 0.402 & 97, 131\\
  Be 73 & 215.28 & -9.42 & 95.7 & 0.2 & 9800 & 16.9 & -1.606 & 9.2 & -42.6 & -0.27 & 24, 25\\
  Be 81 & 34.51 & -2.07 & 48.1 & 0.5 & 3200 & 5.6 & -0.115 & 9.0 & 187.1 & 0.803 & 105, 137\\
  Be 82 & 46.81 & 1.59 & -2.8 & 0.8 & 1000 & 7.4 & 0.027 & 7.85 & 172.3 & 0.791 & 44, 84\\
  Blanco 1 & 15.57 & -79.26 & 5.5 & 0.8 & 200 & 8.0 & -0.236 & 7.7 & 25.6 & 0.05 & 37, 85\\
  Cr 110 & 209.65 & -1.98 & 40.3 & 3.6 & 2100 & 9.9 & -0.071 & 9.15 & -81.9 & -0.402 & 17, 89\\
  Cr 223 & 286.19 & -1.88 & -7.9 & 0.9 & 2900 & 7.7 & -0.095 & 7.56 & -227.8 & -0.996 & 32, 84\\
  Cr 261 & 301.68 & -5.53 & -26.0 & 1.0 & 2200 & 7.1 & -0.211 & 9.9 & -217.1 & -0.954 & 30, 54\\
  Hyades & 180.06 & -22.34 & 39.3 & 2.8 & 0.0 & 8.0 & -0.017 & 8.85 & 33.4 & -0.0010 & 55, 85\\
  IC 2391 & 270.36 & -6.84 & 14.5 & 0.6 & 200 & 8.0 & -0.019 & 7.7 & -215.0 & -0.993 & 12, 85\\
  IC 2488 & 277.83 & -4.42 & -2.7 & 0.0 & 1200 & 7.9 & -0.096 & 8.25 & -230.1 & -1.0 & 36, 93\\
  IC 2602 & 289.6 & -4.91 & 18.1 & 0.9 & 200 & 7.9 & -0.013 & 7.5 & -196.0 & -0.945 & 85, 101\\
  IC 2714 & 292.4 & -1.8 & -13.6 & 0.5 & 1300 & 7.6 & -0.041 & 8.5 & -224.4 & -0.973 & 34, 84\\
  IC 4651 & 340.09 & -7.91 & -31.0 & 0.5 & 800 & 7.3 & -0.109 & 9.4 & -99.9 & -0.371 & 71, 84\\
  IC 4756 & 36.38 & 5.24 & -25.1 & 0.7 & 400 & 7.7 & 0.035 & 8.7 & 118.5 & 0.614 & 61, 84\\
  M 67 & 215.7 & 31.9 & 33.5 & 1.4 & 800 & 8.6 & 0.44 & 9.6 & -91.3 & -0.462 & 84, 108\\
  Mel 66 & 259.56 & -14.24 & 23.0 & 3.0 & 3000 & 9.0 & -0.743 & 9.8 & -198.0 & -0.847 & 45, 60\\
  Mel 71 & 228.95 & 4.5 & 50.7 & 0.4 & 2100 & 9.5 & 0.164 & 9.0 & -130.1 & -0.633 & 70, 84\\
  Mel105 & 292.9 & -2.41 & 0.4 & 0.3 & 2300 & 7.4 & -0.096 & 8.4 & -209.4 & -0.992 & 84, 107\\
  NGC 129 & 120.27 & -2.54 & -37.4 & 0.6 & 1700 & 9.0 & -0.077 & 7.7 & 158.6 & 0.767 & 84, 96\\  
  NGC 188 & 122.84 & 22.38 & -42.4 & 0.0 & 1700 & 8.9 & 0.65 & 9.8 & 130.7 & 0.694 & 50, 112\\
  NGC 436 & 126.11 & -3.91 & -73.8 & 0.6 & 3200 & 10.2 & -0.221 & 7.6 & 108.4 & 0.63 & 84, 96\\
  NGC 457 & 126.64 & -4.38 & -29.8 & 0.5 & 3000 & 10.1 & -0.231 & 7.1 & 151.0 & 0.634 & 84, 96\\
  NGC 581 & 128.05 & -1.8 & -44.2 & 0.7 & 2700 & 9.9 & -0.085 & 7.2 & 133.1 & 0.637 & 84, 96\\
  NGC 654 & 129.08 & -0.36 & -34.2 & 1.2 & 2700 & 9.9 & -0.017 & 7.2 & 140.3 & 0.626 & 84, 96\\
  NGC 663 & 129.47 & -0.94 & -33.1 & 1.7 & 2800 & 10.0 & -0.046 & 6.3 & 140.4 & 0.616 & 84, 96\\
  NGC 884 & 135.05 & -3.58 & -42.6 & 0.6 & 2300 & 9.8 & -0.146 & 7.1 & 115.0 & 0.576 & 84, 116\\
  NGC 1027 & 135.76 & 1.52 & 15.6 & 0.3 & 1000 & 8.8 & 0.027 & 8.4 & 170.8 & 0.636 & 81, 84\\
  NGC 1193 & 146.75 & -12.2 & -82.0 & 9.0 & 4600 & 12.0 & -0.966 & 9.7 & 36.5 & 0.358 & 47, 75\\
  NGC 1528 & 152.06 & 0.26 & -9.7 & 0.5 & 1100 & 9.0 & 0.0050 & 8.6 & 91.0 & 0.417 & 84, 115\\
  NGC 1647 & 180.34 & -16.77 & -7.0 & 0.3 & 400 & 8.4 & -0.128 & 8.6 & -14.9 & -0.005 & 59, 84\\
  NGC 1662 & 187.7 & -21.11 & -13.4 & 0.4 & 400 & 8.4 & -0.138 & 8.4 & -48.1 & -0.12 & 58, 84\\
  NGC 1778 & 168.9 & -2.02 & 4.9 & 2.2 & 1700 & 9.6 & -0.059 & 8.2 & 41.0 & 0.16 & 9, 84\\
  NGC 1817 & 186.16 & -13.1 & 65.3 & 0.5 & 1800 & 9.8 & -0.412 & 8.7 & 33.9 & -0.086 & 56, 84\\
  NGC 1883 & 163.08 & 6.16 & -30.8 & 0.6 & 4800 & 12.7 & 0.517 & 9.0 & 27.0 & 0.183 & 21, 130\\
  NGC 1912 & 172.25 & 0.69 & -45.0 & 0.4 & 1400 & 9.4 & 0.017 & 8.6 & -22.7 & 0.115 & 84, 91\\
  NGC 2099 & 177.63 & 3.09 & 8.3 & 0.2 & 1500 & 9.5 & 0.064 & 8.75 & 8.5 & 0.035 & 66, 84\\
  NGC 2112 & 205.87 & -12.61 & 32.5 & 2.6 & 1000 & 8.9 & -0.217 & 9.3 & -72.6 & -0.384 & 29, 84\\
  NGC 2141 & 198.04 & -5.81 & 24.1 & 0.7 & 3800 & 11.7 & -0.385 & 9.5 & -55.2 & -0.211 & 27, 136\\
  NGC 2158 & 186.63 & 1.78 & 28.0 & 4.0 & 3600 & 11.6 & 0.113 & 9.3 & -7.9 & -0.08 & 26, 113\\
  NGC 2168 & 186.59 & 2.22 & -8.2 & 0.4 & 800 & 8.8 & 0.032 & 8.3 & -44.0 & -0.104 & 84, 120\\
  NGC 2194 & 197.25 & -2.35 & 7.8 & 0.8 & 2800 & 10.7 & -0.113 & 8.8 & -69.2 & -0.222 & 65, 74\\
  NGC 2204 & 226.01 & -16.11 & 92.3 & 2.6 & 4000 & 11.0 & -1.104 & 9.8 & -72.1 & -0.502 & 68, 84\\
  NGC 2243 & 239.48 & -18.01 & 59.8 & 0.6 & 3700 & 10.3 & -1.156 & 9.7 & -132.4 & -0.638 & 14, 84\\
  NGC 2264 & 202.94 & 2.2 & 24.1 & 3.3 & 800 & 8.7 & 0.029 & 6.2 & -74.8 & -0.358 & 79, 92\\
  NGC 2281 & 174.9 & 16.88 & 19.0 & 0.1 & 500 & 8.5 & 0.139 & 8.75 & 28.2 & 0.08 & 2, 84\\
  NGC 2287 & 231.02 & -10.44 & 23.3 & 0.2 & 700 & 8.5 & -0.129 & 8.4 & -158.4 & -0.723 & 84, 115\\
  NGC 2323 & 221.67 & -1.33 & 6.1 & 0.4 & 900 & 8.7 & -0.022 & 8.0 & -154.6 & -0.609 & 84, 115\\
  NGC 2324 & 213.45 & 3.3 & 41.8 & 0.5 & 4200 & 11.7 & 0.24 & 8.8 & -93.7 & -0.376 & 73, 84\\
  NGC 2354 & 238.37 & -6.79 & 33.5 & 1.0 & 1400 & 8.8 & -0.171 & 9.0 & -166.6 & -0.765 & 33, 84\\
  NGC 2355 & 203.39 & 11.8 & 35.0 & 0.4 & 1900 & 9.7 & 0.39 & 9.0 & -64.7 & -0.319 & 4, 84\\
  NGC 2360 & 229.81 & -1.43 & 27.3 & 0.8 & 1100 & 8.8 & -0.029 & 8.7 & -155.5 & -0.696 & 41, 84\\
  NGC 2374 & 228.41 & 1.02 & 29.2 & 0.4 & 1100 & 8.8 & 0.02 & 7.9 & -150.3 & -0.68 & 7, 84\\
  NGC 2420 & 198.11 & 19.63 & 73.6 & 0.6 & 2600 & 10.4 & 0.884 & 9.3 & -4.8 & -0.226 & 84, 115\\
  NGC 2437 & 231.86 & 4.06 & 49.2 & 0.6 & 1500 & 9.0 & 0.107 & 8.4 & -138.8 & -0.697 & 84, 115\\
  NGC 2439 & 246.44 & -4.47 & 66.0 & 0.2 & 4400 & 10.5 & -0.34 & 7.3 & -149.1 & -0.694 & 84, 134\\
  NGC 2447 & 240.04 & 0.14 & 12.1 & 0.6 & 1100 & 8.6 & 0.0020 & 8.6 & -193.4 & -0.808 & 15, 84\\
  NGC 2451 & 252.05 & -6.73 & 16.8 & 0.3 & 200 & 8.1 & -0.022 & 7.8 & -204.3 & -0.938 & 84, 99\\
  NGC 2477 & 253.56 & -5.84 & 7.3 & 1.0 & 1300 & 8.4 & -0.128 & 9.0 & -215.9 & -0.904 & 68, 84\\
  NGC 2489 & 246.71 & -0.77 & 38.2 & 0.3 & 1800 & 8.9 & -0.024 & 8.7 & -178.3 & -0.829 & 84, 98\\
  NGC 2506 & 230.56 & 9.94 & 83.2 & 1.6 & 3200 & 10.3 & 0.546 & 9.25 & -100.1 & -0.593 & 69, 84\\
  NGC 2516 & 273.82 & -15.86 & 23.1 & 0.5 & 400 & 8.0 & -0.098 & 8.2 & -197.1 & -0.962 & 84, 122\\
  NGC 2527 & 246.09 & 1.86 & 39.6 & 0.1 & 600 & 8.3 & 0.019 & 9.0 & -176.2 & -0.885 & 39, 84\\
  NGC 2533 & 247.8 & 1.31 & 35.2 & 0.6 & 2800 & 9.4 & 0.064 & 8.2 & -183.1 & -0.785 & 76, 84\\
  NGC 2539 & 233.71 & 11.11 & 28.9 & 0.7 & 1100 & 8.7 & 0.211 & 8.8 & -161.2 & -0.729 & 31, 84\\
  NGC 2546 & 254.85 & -1.99 & 15.2 & 0.7 & 1000 & 8.3 & -0.034 & 7.0 & -210.7 & -0.929 & 77, 84\\
  NGC 2548 & 227.87 & 15.39 & 7.7 & 0.4 & 700 & 8.5 & 0.192 & 8.6 & -165.9 & -0.675 & 8, 84\\
  NGC 2567 & 249.79 & 2.96 & 36.2 & 0.3 & 1600 & 8.7 & 0.084 & 8.7 & -184.7 & -0.863 & 84, 102\\
  NGC 2627 & 251.58 & 6.65 & 26.0 & 2.1 & 1800 & 8.8 & 0.214 & 9.25 & -196.3 & -0.862 & 1, 84\\
  NGC 2660 & 265.93 & -3.01 & 21.3 & 1.0 & 2800 & 8.6 & -0.145 & 9.0 & -210.0 & -0.922 & 84, 109\\
  NGC 2818 & 261.98 & 8.58 & 20.7 & 1.0 & 1800 & 8.4 & 0.264 & 9.05 & -208.8 & -0.93 & 84, 95\\
  NGC 3114 & 283.33 & -3.84 & -1.7 & 0.3 & 900 & 7.8 & -0.063 & 8.2 & -224.4 & -0.991 & 84, 106\\
  NGC 3293 & 285.86 & 0.07 & -13.6 & 1.4 & 2800 & 7.7 & 0.0040 & 6.9 & -234.4 & -0.997 & 11, 84\\
  NGC 3532 & 289.57 & 1.35 & 4.3 & 1.0 & 500 & 7.8 & 0.012 & 8.3 & -211.4 & -0.96 & 43, 84\\
  NGC 3680 & 286.76 & 16.92 & 1.3 & 0.3 & 1000 & 7.8 & 0.291 & 9.3 & -210.8 & -0.942 & 6, 84\\ 
  NGC 3766 & 294.12 & -0.03 & -16.7 & 0.5 & 2200 & 7.4 & -0.0010 & 7.4 & -224.7 & -0.989 & 84, 87\\
  NGC 3960 & 294.37 & 6.18 & -22.3 & 1.1 & 2100 & 7.4 & 0.225 & 8.95 & -229.4 & -0.98 & 19, 84\\
  NGC 4463 & 300.64 & -2.01 & -12.2 & 0.3 & 1000 & 7.6 & -0.033 & 7.3 & -206.8 & -0.91 & 38, 84\\
  NGC 4755 & 303.21 & 2.5 & -21.2 & 0.9 & 2100 & 7.1 & 0.091 & 7.0 & -210.5 & -0.945 & 84, 110\\
  NGC 5138 & 307.54 & 3.52 & -9.6 & 1.0 & 1400 & 7.2 & 0.087 & 8.5 & -188.2 & -0.877 & 78, 84\\
  NGC 5316 & 310.23 & 0.12 & -15.1 & 0.3 & 1200 & 7.3 & 0.0020 & 8.1 & -186.4 & -0.837 & 84, 94\\
  NGC 5460 & 315.75 & 12.63 & -5.2 & 0.2 & 800 & 7.5 & 0.173 & 8.3 & -158.5 & -0.729 & 10, 84\\
  NGC 5617 & 314.67 & -0.1 & -35.8 & 0.8 & 2000 & 6.7 & -0.0040 & 8.15 & -194.4 & -0.845 & 1, 84\\
  NGC 5662 & 316.94 & 3.39 & -21.3 & 0.3 & 800 & 7.4 & 0.047 & 7.9 & -173.3 & -0.733 & 84, 104\\
  NGC 5822 & 321.57 & 3.59 & 29.3 & 0.8 & 900 & 7.3 & 0.058 & 9.1 & -108.0 & -0.681 & 84, 127\\
  NGC 6067 & 329.74 & -2.2 & -40.0 & 0.6 & 1600 & 6.7 & -0.062 & 7.9 & -148.8 & -0.605 & 3, 84\\
  NGC 6124 & 340.74 & 6.02 & -21.1 & 0.5 & 600 & 7.5 & 0.059 & 8.0 & -89.5 & -0.351 & 84, 94\\
  NGC 6134 & 334.92 & -0.2 & 25.7 & 0.7 & 700 & 7.4 & -0.0020 & 8.95 & -64.5 & -0.46 & 71, 84\\
  NGC 6192 & 340.65 & 2.12 & -7.9 & 0.4 & 1000 & 7.0 & 0.038 & 8.0 & -76.5 & -0.376 & 71, 84\\
  NGC 6208 & 333.76 & -5.76 & -32.5 & 0.3 & 1100 & 7.1 & -0.108 & 9.1 & -125.8 & -0.499 & 78, 84\\
  NGC 6253 & 335.46 & -6.25 & -29.9 & 4.6 & 1700 & 6.5 & -0.185 & 9.5 & -116.7 & -0.508 & 88, 114\\
  NGC 6259 & 342.0 & -1.52 & -34.7 & 0.6 & 1900 & 6.2 & -0.05 & 7.8 & -97.6 & -0.398 & 5, 84\\
  NGC 6268 & 346.05 & 1.3 & -15.1 & 0.1 & 1100 & 6.9 & 0.026 & 8.4 & -62.5 & -0.279 & 35, 84\\
  NGC 6281 & 347.73 & 1.97 & -5.6 & 0.4 & 600 & 7.5 & 0.019 & 8.0 & -46.3 & -0.228 & 42, 84\\
  NGC 6404 & 355.66 & -1.18 & 10.6 & 1.1 & 1700 & 6.3 & -0.036 & 8.7 & 2.2 & -0.097 & 82\\
  NGC 6405 & 356.58 & -0.78 & -8.0 & 0.5 & 500 & 7.5 & -0.0060 & 8.0 & -12.8 & -0.063 & 84, 132\\
  NGC 6425 & 357.94 & -1.61 & -3.5 & 0.3 & 800 & 7.2 & -0.023 & 8.34 & -2.6 & -0.04 & 84, 123\\
  NGC 6475 & 355.86 & -4.5 & -14.8 & 1.3 & 300 & 7.7 & -0.02 & 8.3 & -22.0 & -0.074 & 85, 117\\
  NGC 6520 & 2.88 & -2.84 & -23.5 & 0.4 & 1900 & 6.1 & -0.097 & 8.2 & -2.6 & 0.066 & 28, 84\\
  NGC 6633 & 36.01 & 8.33 & 29.0 & 0.2 & 300 & 7.8 & 0.046 & 9.0 & 170.1 & 0.601 & 62, 84\\
  NGC 6649 & 21.64 & -0.79 & -8.6 & 0.4 & 1600 & 6.6 & -0.021 & 7.7 & 85.4 & 0.45 & 84, 133\\
  NGC 6705 & 27.31 & -2.78 & 35.1 & 1.2 & 2000 & 6.3 & -0.099 & 8.4 & 149.7 & 0.586 & 84, 121\\
  NGC 6709 & 42.12 & 4.72 & -10.2 & 0.6 & 1200 & 7.2 & 0.098 & 8.5 & 150.9 & 0.747 & 84, 119\\
  NGC 6755 & 38.6 & -1.69 & 26.6 & 0.1 & 2100 & 6.5 & -0.063 & 8.4 & 178.5 & 0.771 & 71, 84\\
  NGC 6791 & 69.96 & 10.9 & -57.0 & 2.0 & 3600 & 7.6 & 0.684 & 10.1 & 158.7 & 0.976 & 112, 118\\
  NGC 6802 & 55.33 & 0.92 & 12.4 & 0.6 & 1800 & 7.1 & 0.028 & 8.98 & 208.2 & 0.921 & 63, 137\\
  NGC 6811 & 79.21 & 12.02 & 7.3 & 0.2 & 1000 & 7.9 & 0.213 & 8.8 & 230.4 & 0.977 & 52, 84\\
  NGC 6819 & 73.98 & 8.48 & 4.8 & 0.9 & 2400 & 7.7 & 0.356 & 9.4 & 226.8 & 0.989 & 16, 53\\
  NGC 6939 & 95.9 & 12.3 & -19.0 & 0.2 & 1600 & 8.3 & 0.333 & 9.1 & 204.1 & 0.937 & 81, 86\\
  NGC 7142 & 105.35 & 9.48 & -50.3 & 0.3 & 2300 & 8.9 & 0.386 & 9.84 & 166.9 & 0.855 & 63, 64\\
  NGC 7654 & 112.82 & 0.43 & -33.0 & 0.3 & 1400 & 7.9 & 0.288 & 8.2 & 177.3 & 0.938 & 84, 90\\
  NGC 7789 & 115.53 & -5.39 & -54.9 & 0.1 & 1800 & 8.9 & -0.171 & 9.2 & 150.3 & 0.804 & 51, 83\\ 
  Pleiades & 166.57 & -23.52 & 5.9 & 1.0 & 100 & 8.1 & -0.046 & 7.95 & 50.1 & 0.21 & 13, 85\\
  Praesepe & 205.92 & 32.48 & 34.8 & 0.8 & 200 & 8.1 & 0.092 & 8.86 & -61.4 & -0.363 & 85, 128\\
  Ru 4 & 222.05 & -5.34 & 47.5 & 1.0 & 4700 & 11.9 & -0.439 & 8.9 & -113.2 & -0.448 & 23, 25\\
  Ru 7 & 225.42 & -4.62 & 76.6 & 0.5 & 6000 & 12.9 & -0.482 & 8.9 & -93.9 & -0.44 & 23, 25\\
  Ru 97 & 296.79 & -0.49 & -14.9 & 0.7 & 4100 & 7.2 & -0.035 & 9.0 & -217.9 & -0.997 & 84, 129\\
  Saurer 1 & 214.69 & 7.39 & 104.6 & 0.2 & 11600 & 18.6 & 1.491 & 9.7 & -34.6 & -0.242 & 20, 22\\
  Stock 2 & 133.33 & -1.69 & -17.4 & 0.4 & 300 & 8.2 & -0.0090 & 8.0 & 145.3 & 0.708 & 84, 111\\
  To 2 & 232.83 & -6.88 & 125.0 & 2.9 & 6200 & 12.7 & -0.741 & 9.6 & -63.1 & -0.498 & 46, 72\\
  Tr 2 & 137.38 & -3.97 & -3.2 & 0.4 & 700 & 8.5 & -0.049 & 7.95 & 147.4 & 0.633 & 49, 100\\
  Tr 3 & 138.01 & 4.5 & -8.6 & 1.6 & 700 & 8.5 & 0.054 & 7.83 & 138.9 & 0.625 & 80, 84\\
  Tr 20 & 301.48 & 2.22 & -40.8 & 1.0 & 3300 & 6.9 & 0.128 & 9.11 & -234.1 & -0.991 & 100\\
  Tr 35 & 28.28 & 0.01 & 25.9 & 0.4 & 2100 & 6.2 & 0.0 & 7.4 & 143.7 & 0.607 & 84, 126\\
\enddata
\tablenotetext{a}{$V_S$ and cos$(\psi)$ as given in the text}
\tablerefs{(1) \citealt{ahu05}; (2) \citealt{alc67}; (3) \citealt{an07}; (4) \citealt{ann99}; (5) \citealt{at89}; (6) \citealt{at91}; (7) \citealt{bab85}; (8) \citealt{bn05}; (9) \citealt{bh73}; (10) \citealt{baby95}; (11) \citealt{bau03}; (12) \citealt{rb97}; (13) \citealt{bel98}; (14) \citealt{ber91}; (15) \citealt{bobi05}; (16) \citealt{bra01}; (17) \citealt{brto03}; (18) \citealt{bra06a}; (19) \citealt{bra06b}; (20) \citealt{cb03}; (21) \citealt{car03}; (22) \citealt{car04}; (23) \citealt{car05a}; (24) \citealt{car05b}; (25) \citealt{car07}; (26) \citealt{car02}; (27) \citealt{car01}; (28) \citealt{car05c}; (29) \citealt{car08}; (30) \citealt{carre05}; (31) \citealt{cho03}; (32) \citealt{cla91}; (33) \citealt{cla98}; (34) \citealt{cla94}; (35) \citealt{cla06}; (36) \citealt{cla03}; (37) \citealt{dee85}; (38) \citealt{del07}; (39) \citealt{dod77}; (40) \citealt{dur01}; (41) \citealt{egg68}; (42) \citealt{ff74}; (43) \citealt{fs80}; (44) \citealt{for86}; (45) \citealt{frja93}; (46) \citealt{f02}; (47) \citealt{f89}; (48) \citealt{fri06}; (49) \citealt{fro06}; (50) \citealt{gel08}; (51) \citealt{gim98}; (52) \citealt{glu99}; (53) \citealt{glu93}; (54) \citealt{goz96}; (55) \citealt{han75}; (56) \citealt{haha77}; (57) \citealt{hase04}; (58) \citealt{hass72a}; (59) \citealt{hass72b}; (60) \citealt{haw76}; (61) \citealt{her75}; (62) \citealt{hil58}; (63) \citealt{jh11}; (64) \citealt{jac08}; (65) \citealt{jac11}; (66) \citealt{kal01}; (67) \citealt{kal97}; (68) \citealt{kas97}; (69) \citealt{kim01}; (70) \citealt{kim99}; (71) \citealt{kjfr91}; (72) \citealt{kub92}; (73) \citealt{kye01}; (74) \citealt{kye05}; (75) \citealt{kye08}; (76) \citealt{lin68a}; (77) \citealt{lin68b}; (78) \citealt{lin72}; (79) \citealt{liu89}; (80) \citealt{mabu10}; (81) \citealt{mani07}; (82) \citealt{mag10}; (83) \citealt{mar94}; (84) \citealt{mer08}; (85) \citealt{mer09}; (86) \citealt{mil94}; (87) \citealt{moi97}; (88) \citealt{mon09}; (89) \citealt{pan10}; (90) \citealt{pan01}; (91) \citealt{pan07}; (92) \citealt{par00}; (93) \citealt{ped87a}; (94) \citealt{ped87b}; (95) \citealt{ped89}; (96) \citealt{phja94}; (97) \citealt{phja96}; (98) \citealt{pia07}; (99) \citealt{pla01}; (100) \citealt{pla08}; (101) \citealt{pro96}; (102) \citealt{rapo92}; (103) \citealt{rova98}; (104) \citealt{saca97}; (105) \citealt{sg98}; (106) \citealt{sash91}; (107) \citealt{sag01}; (108) \citealt{san04}; (109) \citealt{san99}; (110) \citealt{san01}; (111) \citealt{sage01}; (112) \citealt{sar99}; (113) \citealt{s95}; (114) \citealt{ses07}; (115) \citealt{sha06}; (116) \citealt{sle02}; (117) \citealt{sno76}; (118) \citealt{ste03}; (119) \citealt{susa99}; (120) \citealt{sube99}; (121) \citealt{sun99};  (122) \citealt{sun02}; (123) \citealt{thst70}; (124) \citealt{tos04}; (125) \citealt{tos98}; (126) \citealt{tur80}; (127) \citealt{twa93}; (128) \citealt{upg79}; (129) \citealt{van76}; (130) \citealt{vil07}; (131) \citealt{vil05}; (132) \citealt{vle74}; (133) \citealt{wala87}; (134) \citealt{whi75}; (135) \citealt{yon12}; (136) \citealt{yon05} ; (137) This work}
\label{sample_params}
\end{deluxetable}

\begin{deluxetable}{l c c c c c c}
\tablewidth{0pt}
\tablecolumns{7}
\tablecaption{Rotational Velocity Solutions}
\tablehead{\colhead{} & \colhead{Average Age} & \colhead{$\sigma_{age}$} & \colhead{V$_{rot}$} & \colhead{$\epsilon_{rot}$} & \colhead{$\sigma_{los}$} & \colhead{Number of} \\ \colhead{Age Range} & \colhead{(Gyr)} & \colhead{(Gyr)} & \colhead{(km s$^{-1}$)} & \colhead{(km s$^{-1}$)} & \colhead{(km s$^{-1}$)} & \colhead{Clusters}}
\startdata
 \textless \, 100 Myr & & & 221 & 3 & 10 & 35\\
 \textgreater \, 1 Gyr & & & 210 & 6 & 26 & 50\\
  \textgreater \, 4 Gyr & & & 187 & 14 & 32 & 17\\
  1 -- 30 Myr & 0.015 & 0.008 & 223 & 3 & 8 & 14\\
  30 -- 100 Myr & 0.074 & 0.021 & 219 & 4 & 11 & 21\\
  100 -- 400 Myr & 0.22 & 0.06 & 220 & 4 & 14 & 25\\
  400 Myr -- 1 Gyr & 0.6 & 0.16 & 224 & 8 & 22 & 24\\
  1 -- 2 Gyr & 1.2 & 0.3 & 221 & 7 & 19 & 20\\
  2 -- 5 Gyr & 2.9 & 0.7 & 213 & 12 & 24 & 17\\
  Over 5 Gyr & 6.7 & 2.0 & 189 & 16 & 33 & 13\\
\enddata
\label{vr_solns}
\end{deluxetable}

\end{document}